%% file: main.tex
\documentclass[aps,pra,twocolumn,showpacs,superscriptaddress,floatfix]{revtex4-2}
\usepackage{amssymb,amsmath}
\usepackage{graphicx}
\usepackage{float}
\usepackage{bm}
\usepackage{dcolumn}
\usepackage{tabularx}
\usepackage{longtable}
\usepackage[dvipsnames,usenames]{color}
\usepackage{gensymb}
\usepackage{enumitem}  %for roman numerals in enumeration etc .... APMCM 29/3/2016.
\usepackage{tikz}
\usepackage{pgfplots} %plot directly in latex
\usepackage[caption=false]{subfig} %for subfloats
\usepackage{placeins}
\usepackage[normalem]{ulem}
\usepackage{hyperref}
\hypersetup{
    colorlinks,
    citecolor=blue,
    filecolor=blue,
    linkcolor=blue,
    urlcolor=magenta
}

\newcommand{\ud}[1]{{\color{black}{#1}}}

\begin{document}
%\title{Novel Polar Phase Induced by Unusual pseudoTriggered Mechanism}
\title{Dielectric softening in the halide double perovskites $A_2$Au$_2X_6$ ($A$: Cs, Rb; $X$: Cl, Br, I) via a strain-mediated pseudotriggered mechanism}

\author{Urmimala Dey}
\affiliation{Centre for Materials Physics, Durham University, South Road, Durham DH1 3LE, United Kingdom}
\affiliation{Luxembourg Institute of Science and Technology (LIST), Avenue des Hauts-Fourneaux 5, L-4362 Esch/Alzette, Luxembourg}

\author{Jordan A.R. Cowell}
\affiliation{School of Physical Sciences, University of Kent, Canterbury, CT2 7NH}

\author{Nicholas C. Bristowe} 
\email{nicholas.bristowe@durham.ac.uk}
\affiliation{Centre for Materials Physics, Durham University, South Road, Durham DH1 3LE, United Kingdom}

\date{\today}
%========================================================================= 

%========================================================================= 

\begin{abstract} 
Halide perovskites have emerged as promising candidates for next generation photovoltaic applications, attracting significant attention in recent years. Through first-principles calculations combined with group-theoretical analyses, we investigate the structural phase diagram of Pb-free Jahn-Teller-active $A_2$Au$_2X_6$ ($A$: Cs, Rb; $X$: Cl, Br, I) double perovskites. Our study identifies a previously unreported ferroelectric phase, where the softening of the polar mode$-$key to ferroelectricity$-$is driven by an unconventional and indirect coupling with improper strains originating from Jahn-Teller distortions. The proposed strain mediated \textit{pseudo}triggered mechanism offers an alternative pathway to enhance the static dielectric constant or even promote (photo-)ferroelectricity, addressing challenges such as defects, excitons, and charge scattering that hinder photovoltaic efficiency. More broadly, this unique mechanism could be extended to oxide double perovskites and opens up a new type of ferroelectric phase transition worthy of future investigation.
\end{abstract}
%========================================================================= 
\maketitle
%=========================================================================
 \section{Introduction}
 
Highly polarizable materials have a wide range of potential applications, but traditional existing applications include capacitors and microwave dielectrics~\cite{Yang2024,Tan2020,Cava2001}.
A large dielectric constant is therefore often sought after during materials design, and a common strategy is to search for materials with a ferroelectric instability~\cite{Buixaderas2004}.
Apart from maximizing the dielectric constant at typical ferroelectric phase transitions, ferroelectrics themselves offer added functionality, including their non-volatile switching for memory devices, but also inherent piezoelectric and pyroelectric applications~\cite{Scott2007,Li2018,Qiu2020,Pandya2018}. 
Ferroelectric design mechanisms include dipolar molecular ordering (molecular ferroelectrics)~\cite{Zhang2020}, second-order Jahn-Teller active cations or stereochemically active lone pair cations (proper ferroelectrics)~\cite{Cohen1992,Benedek2013,Barone2015,Seshadri2001}, and coupling to nonpolar modes (improper or triggered ferroelectrics)~\cite{Kimura2003,Van2008,Benedek2011,Holakovsky1973,Mercy2017}. 

Highly polarizable ferroelectric-photovoltaic (FE-PV) materials have recently seen a surge of interest within the community~\cite{Choi2009,Seidel2011,Yuan2014}. 
The hybrid halide perovskites, such as (CH$_3$NH$_3$)PbI$_3$ (MAPbI$_3$), have been proposed as a contender for next generation solar cells, in part due to their favorable optical light absorption and electron-hole recombination characteristics~\cite{Kojima2009,Hao2014}. 
Both the dipolar MA and the stereochemically active Pb$^{2+}$ cations are thought to contribute to a relatively large static dielectric constant of around 53~\cite{Berger2020}. This property reduces exciton and scattering effects, and enhances defect tolerance. Furthermore, MAPbI$_3$ has been proposed to be ferroelectric~\cite{Kim2015,Rakita2017}, which would allow for bulk photovoltaic effect through the shift current mechanism~\cite{Zheng2015,Tan2016}. 
 
Earth abundant and non-toxic double perovskites, with general formula $A_2BB^\prime X_6$, have been extensively studied in recent years as an alternative to MAPbI$_3$~\cite{Wang2024,Debbichi2018,Kangsabanik2020}. While some match the excellent light absorption of MAPI$_3$, the dielectric response, and hence tolerance to defects and electron-hole recombination, has been found to be significantly lower. This reduction is likely due to the replacement of Pb$^{2+}$ with ions of lower polarizability and/or smaller Born effective charge, the exception being the Bi-based double perovskites~\cite{Volonakis2016,Slavney2016}. \ud{While Bi-based halide double perovskites with large polarizability offer stability and low toxicity, typically indirect band gaps, deep defect states, and the prevalence of antisite defects limit their photovoltaic efficiency~\cite{Li2018Bi,Rahane2024}. Sn-based halide perovskites, by contrast, exhibit direct band gaps close to the Shockley–Queisser limit and high carrier mobility, making them more suitable for high-efficiency solar cells~\cite{Konstantakou2017}. However, their performance is often compromised by the oxidation of Sn(II) to Sn(IV), which introduces deep-level defects and significantly reduces device stability~\cite{Konstantakou2017,Aktas2022}}.
%Inorganic double perovskites also lack the large dipolar response of MA.
%***Seems inorganic Bi double perovskites and Pb single have similar dielectric constants of ~17 (static ionic). Going to Bi and Pb-free double typically reduces further (~13 for most of you Au systems, and ~8 for the Sn-vacancy) (cite: Wilson APL Mater. 7, 010901 (2019), Maughan Chem. Mater. 2018, 30, 3909−3919, Steele ACS Nano 2018, 12, 8081−8090). Excitonic binding energies?***
 
Recently, gold-based halide double perovskites have gained considerable interest because of their good absorption~\cite{Debbichi2018}, semiconducting properties with optimal band gaps in the visible range~\cite{Liu1999,Kojima2000}, and weakly bound excitonic effects~\cite{Giorgi2018}. \ud{The strong covalent Au–I bonding and multivalent nature of Au help suppress common defects such as cation antisites, which are a significant source of performance degradation in other lead-free double perovskite families.} However, Kangsabanik et al. have demonstrated that the nonpolar gold halides are still prone to moderate defect formation, which can adversely affect their PV performance~\cite{Kangsabanik2020,Zhang2021}. In contrast, the presence of a polar phase (if achievable) could potentially overcome these limitations as the bulk photovoltaic effect in noncentrosymmetric materials can generate dissipationless photocurrent of topological origin, potentially enhancing PV efficiency and enabling additional applications in ultrafast devices~\cite{Morimoto2016,Noma2023,Zeng2024,Dey2024}. Furthermore, the polar phase could enable a high static dielectric constant through an ionic response rather than a dipolar one, thereby reducing the impact of defects, excitons, and charge scattering. The enhanced electrostatic screening in the ferroelectric phase is also likely to result in reduced recombinations and extended charge carrier lifetimes~\cite{Yu2024}. 

%We begin by re-examining the structural phase diagram of the previously synthesized $A_2$Au$_2X_6$ ($A$: Cs, Rb; $X$: Cl, Br, I) double perovskites using first-principles calculations. 
The original experimental work by Matsushita et al. demonstrates that Cs$_2$Au$_2$I$_6$ crystallizes in a nonpolar $I4/mmm$ structure (space group 139), featuring corner-shared compressed and elongated AuI$_6$ octahedra arranged in an alternating pattern along the [001] and [110] directions~\cite{Matsushita1997}. Subsequently, various compositions have been synthesized by altering the halogen species and pressure-induced phase transitions were studied~\cite{Denner1979,Kojima1990,Kitagawa1991,Kojima1994,Matsushita1998,Matsushita2007,Scott2012}. Alternatively, substitution of Cs$^+$ by Rb$^+$ at the $A$-site has been shown to stabilize a monoclinic phase with $I2/m$ symmetry ($C2/m$ in different setting, space group 12) in Rb$_2$Au$_2$Br$_6$ and Rb$_2$Au$_2$I$_6$~\cite{Matsushita2005,Strahle1979}. $P-T$ phase diagrams and structural phase transitions under high pressure are also studied in the Rb-family of gold halides~\cite{Kojima1996}. Recently, Morita et al. have investigated various competing metastable phases of Cs$_2$Au$_2X_6$ ($X$: Cl, Br, I) proposing new strategies to stabilize the metastable phases with improved optoelectronic properties~\cite{Morita2024}. However, the prediction or experimental realization of a polar phase in this family of double perovskites is yet to be done. This would not only enhance the understanding of the phase diagram of $A_2$Au$_2X_6$ ($A$: Cs, Rb; $X$: Cl, Br, I) double perovskites but also enable the discovery of previously unexplored properties with potential applications.
 
In this work, we use \textit{ab-initio} calculations to reproduce the experimental ground state phases of Jahn-Teller (JT)-active $A_2$Au$_2X_6$ ($A$: Cs, Rb; $X$: Cl, Br, I) double perovskites. Our group-theoretical analyses, supported by frozen-phonon calculations, describe the structural modes that lead to the ground state structures from the undistorted $A$Au$X_3$ perovskite phase. Moreover, we identify a previously unreported metastable polar phase of Rb$_2$Au$_2$I$_6$ which lies close to the ground state. This novel polar phase with $I4mm$ symmetry is found to arise from an unusual and indirect coupling between structural modes. We show that the JT mode, responsible for the $I4/mmm$ phase of the gold halides, is indirectly coupled to the polar mode through the strain it produces. This indirect coupling turns out to be very large, and much greater than the direct biquadratic (triggered) coupling between the polar and JT modes. \ud{In  triggered ferroelectrics, the softening of the polar mode is triggered by another nonpolar mode already present in the system, or by another mode that is softening simultaneously. In contrast, the pseudotriggered mechanism described in the manuscript can be thought of as a two-step process, where the secondary order parameters (strain modes) produced by the primary mode (JT distortion) indirectly triggers the softening of the polar mode.} Aside from potential applications as photoferroics, and the benefits of highly polarizable PVs, we argue that the rare strain mediated pseudotriggered mechanism is more broadly applicable to JT-active oxide double perovskites and opens up a new-type of ferroelectric phase transition worthy of future investigation. 

\section{Computational Methods}
We performed \textit{ab-initio} calculations based on density functional theory (DFT) as implemented in the Vienna ab initio Simulation Package (VASP)~\cite{VASP1,VASP2}, version 6.3.2. We used PAW pseudopotentials (PBE, potpaw.64)~\cite{pseudo1,pseudo2} with the valence configurations:  $5s^25p^66s^1$ (Cs), $4s^24p^65s^1$ (Rb), $5d^{10}6s^1$ (Au), $3s^23p^5$ (Cl), $4s^24p^5$ (Br),  $5s^25p^5$ (I). PBEsol, a revised Perdew-Burke-Ernzerhof generalised gradient approximation (GGA) for solids, was chosen as the exchange-correlation functional to accurately describe the structural properties of the bulk gold halides~\cite{PBEsol}. We found that a negative pressure of $-0.7$ GPa with PBEsol functional correctly reproduced the structural properties of the halide double perovskites under consideration. Since the strain-phonon coupling in these systems are highly sensitive to the lattice parameters, we used a constant pressure of $-0.7$ GPa for all the PBEsol calculations. Note that this type of negative pressure correction is routinely applied in first-principles studies of ferroelectrics with local and semi-local exchange-correlation functionals~\cite{Zhong1994,Zhong1995,Zhong1995_2,Zhong1996,Sai2000,Wojdel2013}. Convergence tests performed on the 20-atom $I4/mmm$ unit cell of Cs$_2$Au$_2$I$_6$ revealed that a plane wave cutoff of 600 eV and a $k$-mesh grid of $4 \times 4 \times 2$ were sufficient to resolve total energies, stresses, and forces within 0.1 meV/f.u., 0.01 GPa, and 0.1 mev/{\AA}, respectively. An energy convergence criterion of 10$^{-9}$ eV was set for all calculations and full relaxations were performed until the Feymann-Hellman forces on each atom were less than 1 meV/{\AA}. \ud{Spin-orbit coupling (SOC) was included self-consistently within VASP. To verify the robustness of the proposed pseudotriggered mechanism and to obtain more accurate band gap prediction, we employed the screened hybrid Heyd-Scuseria-Ernzerhof (HSE06) exchange-correlation functional~\cite{HSE06_1,HSE06_2}, which incorporates a fraction of exact Hartree-Fock exchange. Structural relaxations were not performed with either SOC or HSE06 functional. No pressure correction was applied in the HSE06 hybrid calculations.}  We utilized the density functional perturbation theory (DFPT) method implemented in VASP to compute the phonon dispersions \ud{and dielectric constants}~\cite{DFPT1,dfpt2}. PHONOPY was employed as the post-processing tool~\cite{phonopy}. A $2 \times 2 \times 2$ phonon supercell was used for the cubic $A$Au$X_3$ perovskites and $\Gamma$-point phonon calculations were performed to obtain the structural instabilities in the $I4/mmm$ phase of the $A_2$Au$_2$$X_6$ double perovskites ($A$: Cs, Rb; $X$: Cl, Br, I). Macroscopic polarization was computed using the Berry phase approach~\cite{Vanderbilt1,Vanderbilt2} implemented within VASP. Web-based ISOTROPY software suite~\cite{isotropy} was utilized for symmetry mode analyses. In particular, we used FINDSYM~\cite{findsym1,findsym2} to determine the symmetries of the VASP output structures and ISODISTORT~\cite{Campbell1,isodistort2} to obtain mode details. INVARIANTS~\cite{invariants1,invariants2} was utilized to find the invariant polynomials in the Landau free energy expansion around the cubic parent structure. Crystal structures were visualized using VESTA~\cite{vesta}.

\section{Results and Discussion}
\begin{figure}[h]
\includegraphics[scale=0.52]{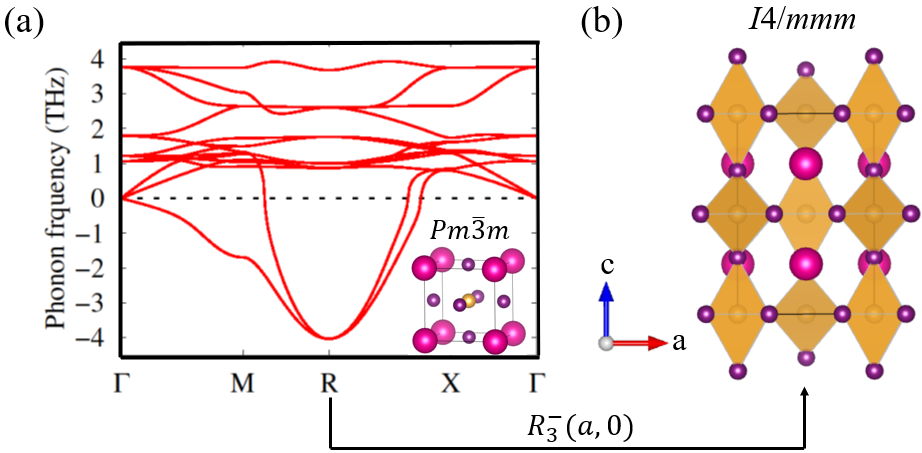}
\caption{(a) Phonon spectrum of the high-symmetry cubic perovskie phase of RbAuI$_3$ as a representative of the gold halide perovskites $A$Au$X_3$ ($A$: Cs, Rb; $X$: Cl, Br, I). (b) Condensation of the most unstable phonon mode with $R^-_3$ symmetry leads to a tetragonal $I4/mmm$ phase with alternating elongated and compressed octahedra. The magenta, gold, and purple spheres represent the Rb, Au, and I atoms, respectively.} 
\label{struct}
\end{figure}

To reproduce the structural phase diagram of the $A_2\rm{Au}_2$$X_6$ ($A$: Cs, Rb; $X$: Cl, Br, I) gold halides, we begin our first-principles investigation with the optimized high-symmetry perovskite structures. Phonon calculations for the simple cubic $A$Au$X_3$ ($A$: Cs, Rb; $X$: Cl, Br, I) perovskites reveal the presence of large instabilies at the zone-boundary \textit{R}-point, located at $\mathbf{k} = (\frac{1}{2}, \frac{1}{2}, \frac{1}{2})$, as shown in Fig.~\ref{struct}(a).
%where $a_c$ is the lattice parameter of the cubic structure. 
Condensation of the lowest frequency phonon mode with $R^-_3$ character leads to a reduction of symmetry from $Pm\bar{3}m$ to $I4/mmm$. In the tetragonal $I4/mmm$ phase, the primary JT distortion $R^-_3(a,0)$ is accompanied by a secondary breathing mode $R^-_2(a)$. The breathing distortion gives rise to alternating corner-sharing elongated and compressed octahedra occupied by $\rm{Au}^{3+}$ and $\rm{Au}^{+}$ ions, respectively, making $A_2\rm{Au}^{+}\rm{Au}^{3+}$$X_6$ an effective double perovskite system with mixed-valency (see Fig.~\ref{struct}(b)). The associated charge disproportionation gives rise to filled Au-$5d$ orbitals at the square-planar and octahedral sites, resulting in nonmagnetic behavior consistent with the previous magnetic susceptibility measurements~\cite{Elliott1934}. 

The JT distortion in the $I4/mmm$ structure additionally produces hydrostatic and tetragonal strains via a linear-quadratic coupling in the Landau free energy expansion. 
\begin{equation}
    \mathcal{F}^{I4/mmm} = \alpha_1 Q_{\Gamma^+_1} Q^2_{R^-_3} + \alpha_2 Q_{\Gamma^+_3} Q^2_{R^-_3}
    \label{eq1}
\end{equation}
where $\alpha_1$ and $\alpha_2$ are the coupling coefficients in the $I4/mmm$ phase. $\Gamma^+_1(a)$ and $\Gamma^+_3(a,0)$ are the irreps associated with the hydrostatic and tetragonal strains, respectively. The $\Gamma^+_1$ strain mode acts as negative hydrostatic pressure, leading to an expansion of the lattice parameters. In contrast, the tetragonal $\Gamma^+_3$ mode reduces the in-plane lattice parameters while expanding the cell along the out-of-plane direction. It should be noted that the strain modes are not the primary order parameters and arise due to the linear-quadratic couplings with the JT distortion in Eq.~(\ref{eq1}). Hence, the 
$I4/mmm$ phase of the studied gold halide can be classified as an improper ferroelastic phase. 

%Goldschmidt tolerance factors and the amplitudes of the structural modes present in the fully relaxed $I4/mmm$ structure of the $A_2\rm{Au}_2X_6$ ($A$: Cs, Rb, K; $X$: Cl, Br, I) gold halide double perovskites are listed in Table~\ref{I4mmm}. We find that smaller tolerance factor leads to a decrease in the overall mode amplitude.
%\begin{table}[ht!]
%  \setlength{\tabcolsep}{6.0pt}
%  \caption{Amplitudes of the structural modes present in the fully relaxed $I4/mmm$ structure of $A_2\rm{Au}_2X_6$ ($A$: Cs, Rb, K; $X$: Cl, Br, I).}
%  \centering
%  \begin{tabular}{|c |c |c | c| c| c|}
%    \hline
%    \hline\rule{0pt}{1.0\normalbaselineskip}
%    \multicolumn{2}{c|}{$A_2\rm{Au}_2X_6$}&\multicolumn{2}{c|}{Strain modes}&\multicolumn{2}{c|}{Displacive modes}\\\cline{1-6}    \rule{0pt}{1.2\normalbaselineskip}
%     $A$& $X$ & $\Gamma^+_1(a)$&$\Gamma^+_3(a, 0)$&$R^-_3(a, 0)$&$R^-_2(a)$   \\ 
%     \hline  
%     Cs&Cl& 0.08641&  0.04244&  0.85730 &  0.12504 \\
%     Cs&Br&  0.06959&   0.04282&  0.78887 &  0.07213 \\
%     Cs&I&  0.06459&  0.03899 &  0.79584 &   0.05123\\
%     Rb&Cl&  0.07554&   0.04060&  0.77256 &  0.10227 \\
%     Rb&Br&   0.06262&   0.03957 &   0.72101 &   0.06214  \\
%     Rb&I& 0.05954&  0.03708& 0.74766 & 0.04542 \\   
%    \hline 
%    \hline
%\end{tabular}
%  \label{I4mmm}
%\end{table} 

Phonon calculations carried out for the relaxed $I4/mmm$ structure of Cs$_2$Au$_2X_6$ show that the $I4/mmm$ phase is dynamically stable for Cs$_2$Au$_2X_6$ ($X$: Cl, Br, I), consistent with previous experimental findings that identified the $I4/mmm$ phase as the ground state of Cs$_2$Au$_2X_6$~\cite{Matsushita1997,Denner1979,Kojima1990,Kitagawa1991,Kojima1994,Matsushita2007,Scott2012}. We find that $I4/mmm$ is also the ground state of Rb$_2$Au$_2$Cl$_6$.

\begin{figure}[h]
\includegraphics[scale=0.505]{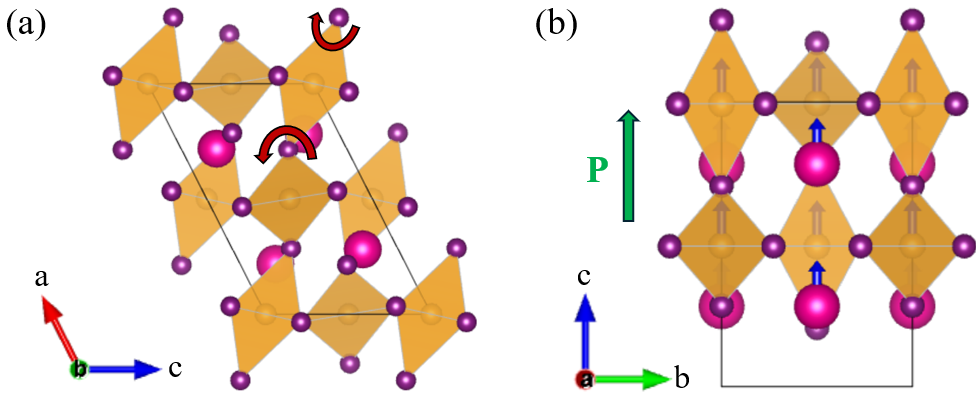}
\caption{(a) Crystal structure of the ground state $C2/m$ phase of Rb$_2$Au$_2$I$_6$. The red arrows indicate the directions of octahedral tilts. (b) The metastable $I4mm$ polar structure of Rb$_2$Au$_2$I$_6$. Polar displacements of the cations and the direction of net out-of-plane polarization ($\mathbf{P}$) are denoted by blue and green arrows, respectively. } 
\label{struct_2}
\end{figure}

%In order to find the ground state structure of , we condense in the unstable phonon eigenmodes at $\Gamma$ and compare the total energies obtained from our first-principles calculations. 
%The resulting In case of Rb-compounds, the $I2/m$ structure contributes to the largest energy lowering from the $I4/mmm$, which arises due to a primary tilt mode transforming as the $R^-_5$ irrep, as seen from Table~2. 
However, as we move towards the gold halides with smaller tolerance factors, the $I4/mmm$ structure no longer remains the ground state and an octahedral tilt mode comes into play which leads to a further reduction of symmetry to a monoclinic $C2/m$ phase in Rb$_2$Au$_2$Br$_6$ and Rb$_2$Au$_2$I$_6$, as observed in previous experiments~\cite{Matsushita2005,Strahle1979}. The $R^-_5$ tilt distortion causes an antiphase rotation of the Au$X_6$ octahedra around the two in-plane axes, as shown in Fig.~\ref{struct_2}(a), and is accompanied by a secondary antipolar $R^-_4$ mode. 
%This $C2/m$ phase has been observed in previous experiments on Rb$_2$Au$_2$I$_6$. Our DFT calculations show that $I2/m$ phase is also the ground state structure of Rb$_2$Au$_2$Br$_6$. 
Dynamical stability of this tilted phase is confirmed from our phonon calculations. Full structural details can be found in the Supplemental Material (SM)~\cite{supp}.
\begin{table*}[ht!]
  \setlength{\tabcolsep}{6.0pt}
  \caption{Frequency of phonon instabilities present in the fully relaxed $I4/mmm$ structure of Rb$_2\rm{Au}_2$I$_6$. The resulting phases, energies of the metastable phases relative to the $I4/mmm$ phase per 20-atom unit cell ($E_{\rm{rel}}$), and the irreducible representations (irreps) of the primary and secondary order parameters (OPs) are also given. The irrep labels are with respect to the cubic parent phase. As seen, the previously unreported $I4mm$ polar phase lies close to the $C2/m$  ground state of Rb$_2\rm{Au}_2$I$_6$.}
  \centering
  \begin{tabular}{|c |c |c | c| c|}
    \hline
    \hline\rule{0pt}{1.0\normalbaselineskip}
%    \multicolumn{2}{c|}{$A_2\rm{Au}_2X_6$}&\multicolumn{2}{c|}{Strain modes}&\multicolumn{2}{c|}{Displacive modes}\\\cline{1-6}    \rule{0pt}{1.2\normalbaselineskip}
     Phonon freq. (cm$^{-1}$)&Resulting phase& $E_{\rm{rel}}$ (meV) & Irreps of primary OPs& Irreps of secondary OPs  \\ 
     \hline \rule{0pt}{1.0\normalbaselineskip}
     13.83$i$ & $I4mm$ &-11.83& $\Gamma^-_4(0, 0, a)$, $R^-_3(a, 0)$ & $\Gamma^+_1(a)$, $\Gamma^+_3(a, 0)$,  \\ \rule{0pt}{1.0\normalbaselineskip}
     &&&&$R^+_5(a, 0 , 0)$, $R^-_2(a)$ \\ \hline\rule{0pt}{1.0\normalbaselineskip} 
     8.70$i$ & $Immm$ & -5.48& $R^-_4(a, 0, 0)$, $R^-_3(a, 0)$&$\Gamma^+_1(a)$, $\Gamma^+_3(a, 0)$, \\  \rule{0pt}{1.0\normalbaselineskip}
      &&&&$\Gamma^+_5(a, 0 , 0)$, $R^-_2(a)$ \\ \hline \rule{0pt}{1.0\normalbaselineskip}
     5.48$i$ & $C2/m$ &-148.18&$R^-_5(a, 0, a)$, $R^-_3(a,1.732a)$&$\Gamma^+_1(a)$, $\Gamma^+_3(a, 1.732a)$, $\Gamma^+_5(a, b , -a)$, \\ \rule{0pt}{1.0\normalbaselineskip}
      &&&&$R^-_2(a)$, $R^-_4(a, b, -a)$ \\ \hline\rule{0pt}{1.0\normalbaselineskip}
     3.56$i$ & $P4_2/nmc$ & -0.14 &$M^-_3(a; 0; 0)$, $R^-_3(a, 0)$ &$\Gamma^+_1(a)$, $\Gamma^+_3(a, 0)$, $R^-_2(a)$\\      
     \hline  
    \hline 
\end{tabular}
\label{RbI}
\end{table*}

Interestingly, while the $C2/m$ phase is the ground state of Rb$_2$Au$_2$I$_6$, our $\Gamma$-phonon calculations for the $I4/mmm$ phase reveals the presence of three additional structural instabilities, which when condensed result in the metastable $I4mm$, $Immm$, and $P4_2/nmc$ structures, as listed in Table~\ref{RbI}. In this work, we focus on the polar $I4mm$ phase which lies just 68.17 meV per 10-atom formula unit 
%(1 f.u. of $A_2\rm{Au}_2X_6$ has 10 atoms) 
above the ground state and possesses a spontaneous polarization of $\sim$ 5.03 $\mu$C/cm$^2$ along the out-of-plane direction (see Fig.~\ref{struct_2}(b)). In this polar phase, the $\Gamma^-_4$ polar mode is the primary distortion that drives the $I4/mmm$ to $I4mm$ phase transition, and appears with the secondary $R^+_5$ antipolar distortion. %Below we demonstrate how the $\Gamma^-_4$ polar mode in the $I4mm$ phase arises due to an unusual coupling with the strain modes resulting from the JT distortion. 
Fig.~S3 of the SM shows the insulating nature of the $I4mm$ structure~\cite{supp}. %One should note that semi-local exchange-correlation functionals like PBEsol are known to underestimate the band gaps and more expensive hybrid functionals have been shown to significantly improve the band gaps of this family of compounds~\cite{Debbichi2018,Kangsabanik2020,Zhang2021}.
\ud{Since semi-local exchange-correlation functionals such as PBEsol are known to underestimate band gaps, we employ the more accurate but computationally demanding HSE06 screened hybrid functional to calculate the density of states of the PBEsol-optimized $I4mm$ structure. We find that HSE06 increases the band gap to approximately 0.9~eV, which is close to the experimentally reported value of $\sim$1.11~eV for the tilted $C2/m$ ground state~\cite{Yu2024RbI}, consistent with the established understanding that octahedral tilts tend to widen the band gap in halide perovskites~\cite{Lee2016}. Total energy calculations using HSE06 further reveal that the polar $I4mm$ structure remains a metastable phase, resulting in an energy lowering of approximately 4.20 meV (compared to $\sim$5.92 meV with PBEsol) per 10-atom formula unit relative to the $I4/mmm$ phase.}

%It is intriguing that the polar mode is induced by a reduction in the tolerance factor and an increase in hydrostatic strain (equivalent to negative hydrostatic pressure), contrary to the typical behavior observed in perovskites~\cite{Benedek2013,Samara1975,Kozlenko2011,Guennou2011}. This raises an important question on the origin of polarization in this phase.

\begin{figure}[h]
\includegraphics[scale=0.28]{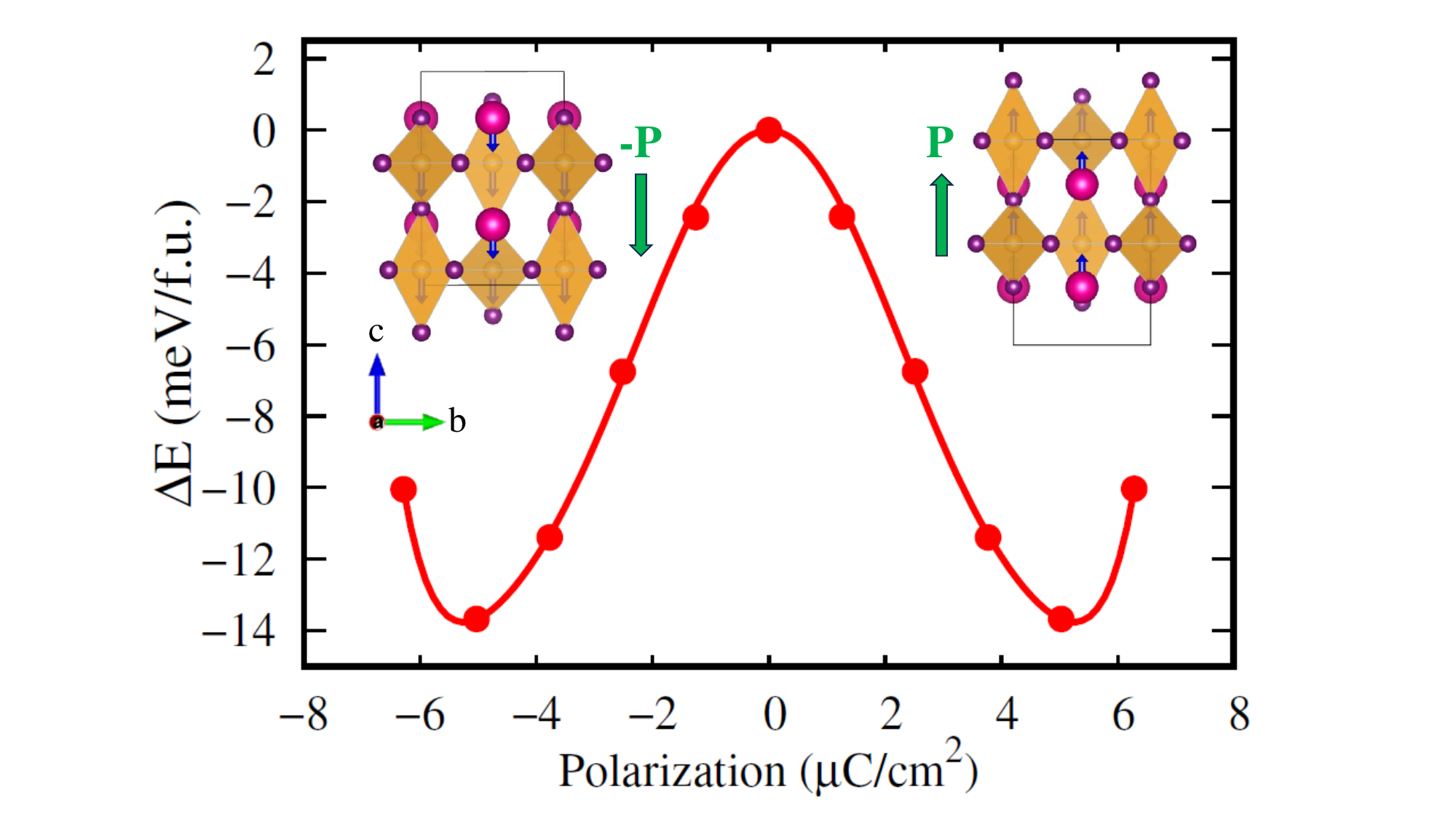}
\caption{\ud{Polarization switching path in the metastable $I4mm$ phase of Rb$_2$Au$_2$I$_6$.}} 
\label{double_well}
\end{figure}
%On the other hand, decrease in the $A$-site cation size from $\rm{Rb}^+$ to $\rm{K}^+$ leads to further reduction in symmetry and as the tolerance factor becomes sufficiently low, a monoclinic $P2_1/n$ phase emerges as the ground state. Note that while the $P2_1/n$ phase of K$_2$Au$_2$I$_6$ has been synthesized experimentally, it is studied in detail. Our group-theoretical analyses reveal that in addition to the modes present in the $I2/m$ structure, the $P2_1/n$ contains two more complex tilt distortions which transform as the $X^+_5$ and $M^+_5$ irreps of the cubic parent structure. Lattice parameters and mode amplitudes of the DFT-relaxed ground state structures are given in Table 3.

%It is interesting to note that all the gold halides in Table~2 exhibit a polar instability in the $I4/mmm$ phase. The resulting polar structure, with $I4mm$ symmetry, reduces energy relative to the high-symmetry tetragonal phase. % 
%This previously unreported polar .0structure lies just above the ground state and, therefore, merits further investigation. Below, we demonstrate how the polar mode arises due to an unusual coupling with strain modes resulting from the Jahn-Teller distortion.
\ud{Fig.~\ref{double_well} shows the polarization double well calculated for the metastable $I4mm$ structure of Rb$_2$Au$_2$I$_6$. The height of the energy barrier, which provides an upper bound for the polarization switching barrier, is approximately 14 meV per 10-atom formula unit, comparable to or even smaller than the values reported for other halide perovskites~\cite{Li2023,Zhou2015} In agreement with experimental observations of polarization switching in halide perovskites~\cite{Coll2015,Wang2017,Zhang2020_2}, this suggests that the polarization in the $I4mm$ phase is likely switchable under an applied electric field.} 

\begin{figure}[h]
\includegraphics[scale=0.7]{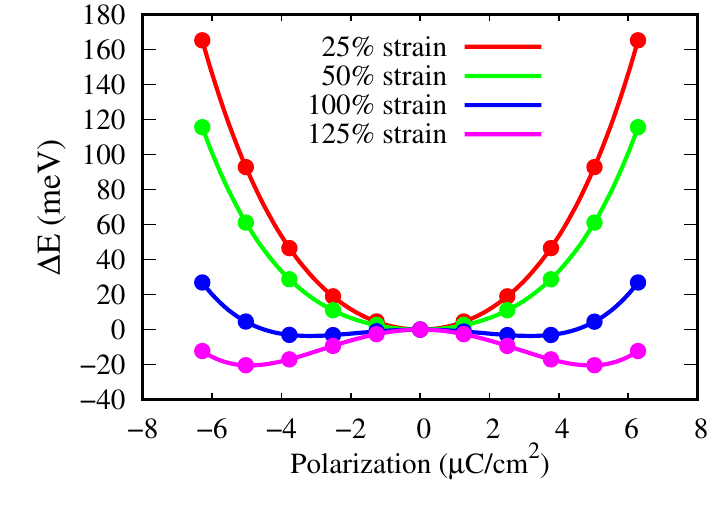}
    \caption{Energy landscape as a function of the polar mode amplitude at different percentage values of the strain modes (both hydrostatic and tetragonal) produced by the JT distortion in the $I4/mmm$ phase.} 
\label{polar_mode}
\end{figure}

To understand the origin of this polar distortion, we plot the energy landscape as a function of the polar mode amplitude at different percentage values of the strain modes (both hydrostatic and tetragonal) produced by the JT distortion in the $I4/mmm$ phase. %We consider both the  strain modes and set the secondary breathing and antipolar modes to zero. 
Fig.~\ref{polar_mode} reveals the transition from a double well to a single well potential as we reduce the strain mode amplitudes from the maximum values produced in the $I4/mmm$ phase. On the other hand, increasing the strain percentage beyond the maximum values results in a deeper well with a larger polarization, highlighting the direct influence of the strain modes on the softening of the polar mode. Note that the polarization magnitude and the depth of the potential well in Fig.~\ref{polar_mode} with 100\% strain are slightly lower than those in the relaxed $I4mm$ phase, as these calculations exclude the contributions from secondary breathing and antipolar modes. \ud{As demonstrated in Fig. S5, we find that the transition from a double well to a single well potential with reduced strain amplitudes is independent of the exchange-correlation functional and remains robust upon inclusion of SOC~\cite{supp}.}

The results in Fig.~\ref{polar_mode} can be framed in the light of an INVARIANTS analysis. Landau free energy expansion around the cubic phase shows the presence of linear-quadratic couplings between both types of strain modes and the polar distortion in the $I4mm$ phase, in addition to the linear-quadratic terms between the strains and the JT distortion.
\begin{equation}
\begin{split}
    \mathcal{F}^{I4mm} &= \gamma_{1}Q^{}_{\Gamma^+_{1}}Q^2_{R^-_{3}} + \gamma_{2}Q^{}_{\Gamma^+_{3}}Q^2_{R^-_{3}}\\
    &+\gamma_{3}Q^{}_{\Gamma^+_{1}}Q^2_{\Gamma^-_{4}}+\gamma_{4}Q^{}_{\Gamma^+_{3}}Q^2_{\Gamma^-_{4}}, 
    \end{split}   
\label{invariants}
\end{equation}
where $\gamma_1$, $\gamma_2$, $\gamma_3$, and $\gamma_4$ are the coupling coefficients in the $I4mm$ phase. 

As in the $I4/mmm$ phase, the JT mode induces the hydrostatic and tetragonal strain modes via the first two terms in Eq.~(\ref{invariants}) (improper ferroelastic). We find that the magnitudes of the strain modes in the relaxed $I4/mmm$ and $I4mm$ structures are very similar (see SM). These strain modes, in turn, renormalize the polar mode instability via the last two linear-quadratic terms, as evident from Fig.~\ref{polar_mode}. Thus, our first-principles calculations suggest that the coupling coefficients in Eq.~(\ref{invariants}) are cooperative and contribute to energy lowering.

\begin{figure}[h]
\includegraphics[scale=0.35]{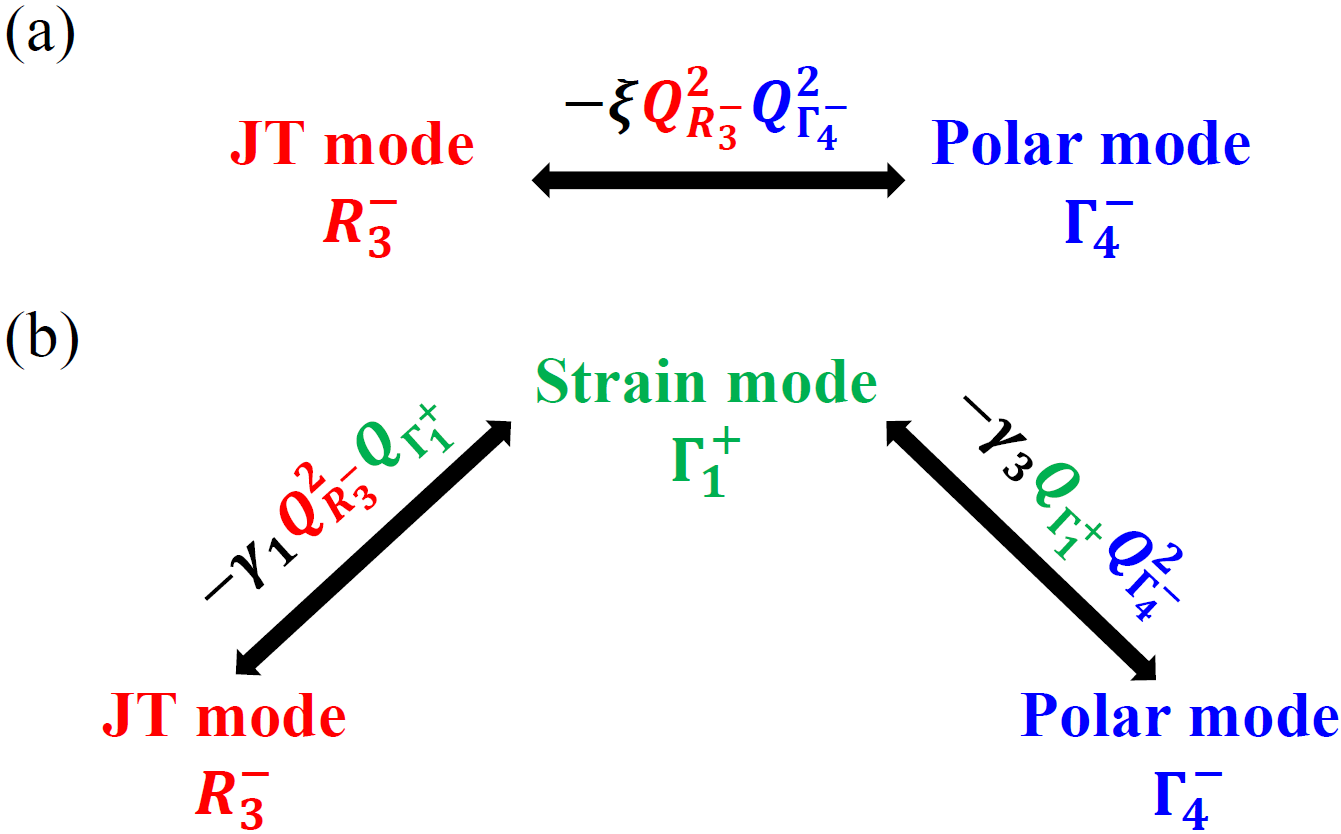}
    \caption{\ud{Schematic diagram illustrating (a) the conventional triggered mechanism and (b) the novel pseudotriggered mechanism. In the conventional triggered mechanism, a cooperative biquadratic coupling exists directly between the two primary order parameters (JT and polar distortions) in the free energy expansion. In contrast, the pseudotriggered mechanism involves an indirect coupling between the two primary modes, mediated by a secondary order parameter (strain) through individual cooperative linear-quadratic couplings, leading to an effective biquadratic coupling between the primary modes—hence the term \textit{pseudo}triggered.}} 
\label{pseudo}
\end{figure}

\begin{figure}[h]
\includegraphics[scale=1.35]{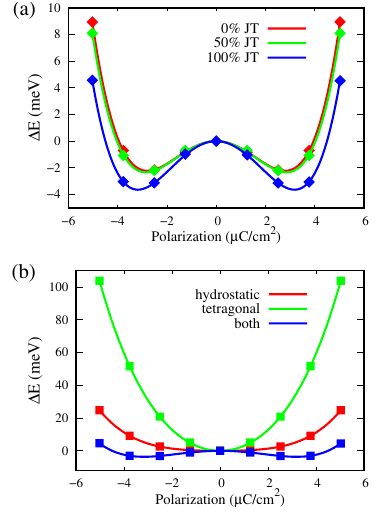}
    \caption{(a) Potential energy landscape as a function of the polar mode amplitude at different percentage size of the JT distortion. Strain mode magnitudes are fixed at their maximum values corresponding to the relaxed $I4/mmm$ structure. (b) Energy landscape in presence of only the hydrostatic strain, tetragonal strain, and both the strains. Here, the magnitudes of the strain and JT modes are fixed at the maximum values.} 
\label{JT}
\end{figure}

%The calculations suggest JT is primary, and via first term in equation above, induces strain (improper ferroelastic). Quantify strain from Pm3-m to I4/mmm. This strain can then renormalise the polar mode instability via the second term. Direct bi-quadratic (no strain), i.e. triggered, is not responsible (see figure). However, an effective bi-quadratic that also includes strain, can be thought of being responsible, i.e. pseudotriggered mediated via strain.
A triggered phase transition occurs when the softening of a structural mode is triggered by another mode already present in the system, or by another mode that is softening simultaneously. The triggered mechanism can be explained by the presence of a cooperative biquadratic coupling term between the two order parameters in the free energy expansion, and has been shown to be the driving mechanism for the metal-insulator transition in rare-earth nickelates~\cite{Mercy2017,Javier2021}. To investigate the possibility of a triggered softening of the polar mode in $\rm{Rb}_2{Au}_2{I}_6$ via a direct cooperative biquadratic coupling between the JT mode and the polar distortion, \ud{as illustrated in Fig.~\ref{pseudo}(a),} we compute the energy landscape as a function of the polar mode amplitude by varying the size of the JT distortion, as shown in Fig.~\ref{JT}(a). Strain mode magnitudes are fixed at their maximum values corresponding to the relaxed $I4/mmm$ structure. From Fig.~\ref{JT}(a), we find that the JT distortion alone has negligible influence on the softening of the polar mode when the strain values are kept fixed. This implies that the strains produced by the JT mode contribute to the lowering of the energy considerably more than the JT distortion itself. Furthermore, we observe a polarization double well even in absence of the JT mode, confirming that the conventional triggered mechanism cannot explain the origin of the polar distortion in this previously unreported $I4mm$ phase.

However, it is important to note that there are linear-quadratic couplings between the strain modes and the JT distortions, as well as between the strain modes and the polar mode (see Eq.~(\ref{invariants})). Our results shown in Fig.~\ref{polar_mode} suggest that the signs of $\gamma_1$ and $\gamma_3$ are the same, and the signs of $\gamma_2$ and $\gamma_4$ are the same, such that the same type of strain favors both the JT mode and the polar mode. Consequently, one could argue that the strains are effectively subsumed within the JT distortion, leading to an effective biquadratic coupling between the $\Gamma^-_4$ polar mode and the $R^-_3$ JT mode ($\approx \lambda Q^2_{\Gamma^-_{4}}Q^2_{R^-_{3}}$). Based on our findings, we conclude that the emergence of the soft polar mode in the $I4mm$ phase of $\rm{Rb}_2{Au}_2{I}_6$ can be attributed to an unconventional \textit{pseudo}triggered mechanism \ud{shown in Fig.~\ref{pseudo}(b)}. To the best of our knowledge, this strain mediated pseudotriggered mechanism has not been reported in literature and warrants future investigation.
%\begin{equation}
% \gamma_{1}Q^{}_{\Gamma^+_{1}}Q^2_{R^-_{3}} + \gamma_{2}Q^{}_{\Gamma^+_{3}}Q^2_{R^-_{3}} + \gamma_{3}Q^{}_{\Gamma^+_{1}}Q^2_{\Gamma^-_{4}}+\gamma_{4}Q^{}_{\Gamma^+_{3}}Q^2_{\Gamma^-_{4}} \approx \lambda Q^2_{\Gamma^-_{4}}Q^2_{R^-_{3}}
%\label{pseudo}
%\end{equation}
%--------------------------------------
It is important to note that while both the strain modes contribute to the energy lowering, in Fig.~\ref{JT}(b), we show that it is the hydrostatic strain that drives the polar phase transition.

Since the strain modes responsible for the polar mode softening are byproducts of the JT distortion, one might expect a higher magnitude of $R^-_3$ mode to result in larger softening. However, as shown in Table~S1 in the SM, Cs-family of compounds with larger $R^-_3$ mode amplitude~\cite{supp} do not exhibit any polar softening. This behavior can be attributed to the increased steepness of the polarization potential wells (large curvature) in high-tolerance-factor halide double perovskites. It suggests that to stabilize the 
$I4mm$ phase one might require a combination of small and large cations at the $A$-site (e.g., both Cs$^+$ and K$^+$), thereby increasing the JT distortion without excessively steepening the potential well. Another possible way is to apply external pressure, which could alter the order of the phonon instabilities in the $I4/mmm$ phase, and remains a topic for future study. Alternatively, epitaxial strain can be used to suppress the octahedral tilts~\cite{Herklotz2016,Park2024}, thereby stabilizing the $I4mm$ phase compared to the tilted $C2/m$ phase in Rb$_2$Au$_2$I$_6$. This could also be achieved by interfacing the Rb$_2$Au$_2$I$_6$ films with an untilted perovskite substrate (since tilts are corner sharing).

More broadly, we propose that the strain-mediated pseudotriggered mechanism explored in our work is applicable to JT-active oxide double perovskites. Examples include  Ba$_2Ln$MoO$_6$ ($Ln$: Nd, Sm, Eu, Gd, Dy, Y, Er, and Yb)~\cite{Cussen2006m}, Ba$_2$FeReO$_6$~\cite{Ferreira2006}, Sr$_2M$ReO$_6$ ($M$: Mg, Cr, Fe, Co, Ni, Zn)~\cite{Kato2004}, $A_2$Cu$B'$O$_6$ ($A$: Ba, Sr; $B'$: W, Te)~\cite{Iwanaga1999}, where the polar softening could also appear via the strains produced by the JT distortion.

%This discovery opens up a novel pathway to enhance the ionic dielectric constant in Pb-free double inorganic perovskites. Additionally, hybrid structures could further amplify dipolar interactions, which is particularly relevant for minimizing the effects of defects, excitons, and charge scattering that can impede photovoltaic efficiency.
 
 %Other halide and oxide systems with the correct symmetry, and supporting calculations. 
 
 %Previous studies of Jahn-Teller coupled to polar mode - trilinear mechanism in vanadates, strained oxides, and hybrid MOF. E-field control of JT. 
 
 %Compare mechanism with known Triggered. Triggered is rare, but can display a variety of types of phase transition. 
 
 %What kind of ferroelectric phase transition might we expect, and how might the dielectric constant vary with T? Strain mediated pseudotriggered mechanism worthy of future investigation.

\section{Conclusions and Outlook}
We have performed first-principles calculations guided by group-theoretical analyses to study the phase diagram of the gold halide double perovskites with the general formula $A_2\rm{Au}_2$$X_6$ ($A$: Cs, Rb; $X$: Cl, Br, I). Our DFT calculations reproduce the $I4/mmm$ ground state of  Cs$_2\rm{Au}_2$$X_6$ ($X$: Cl, Br, I) and Rb$_2$Au$_2$Cl$_6$, where the primary JT distortion induces hydrostatic and tetragonal strains via a linear-quadratic coupling. As the tolerance factor decreases, we observe the emergence of a tilted monoclinic phase in the $C2/m$ space group, consistent with previous experimental findings. Notably, we have identified a previously unreported metastable polar phase of Rb$_2\rm{Au}_2$I$_6$ with $I4mm$ symmetry which lies close to the $C2/m$ ground state. We demonstrate that the strain modes generated by the JT distortion leads to the softening of the out-of-plane polar mode in the $I4mm$ structure, while the direct effect of the JT mode on the polar softening is negligible. Based on our group theory-led DFT calculations, we thus put forward a novel strain mediated pseudotriggered mechanism as the driving force behind the appearance of the polar $I4mm$ phase in Rb$_2\rm{Au}_2$I$_6$. This polar phase, if stabilized, could have potential applications in Pb-free solar cells and ultrafast devices, and therefore, merits further investigation. Additionally, the out-of-plane polarization makes this phase particularly promising for thin-film based device applications. \ud{Moreover, polar halide perovskites have been reported to exhibit larger shift current magnitudes than oxide perovskites, despite having significantly lower polarization values~\cite{Tan2016,Kim2020}. Furthermore, the dielectric constant is expected to \textit{diverge} in the nonpolar $I4/mmm$ phase near the transition to the polar phase upon cooling, suggesting strong dielectric tunability and promising potential for capacitor and dielectric storage applications.} Thus, our discovery of the novel pseudotriggered mechanism is not only of fundamental interest but also opens up new possibilities for tailoring polar phase transitions in JT-active halide and oxide double perovskites with numerous potential applications. Furthermore, the pseudotriggered coupling between the polar mode and the JT mode could enable electric field control of the JT distortion, offering a versatile approach to modulate or activate new electronic properties in JT-active bulk perovskites~\cite{Nick2016}.

\ud{Halide perovskites are known for their strong anharmonicity, which can significantly influence structural and electronic properties at finite temperatures, including lattice expansion and magnitude of phonon softening~\cite{Marronnier2017,Klarbring2020,Seidel2011,Wang2025}. While such temperature-induced effects may impact the relative stability of the metastable polar $I4mm$ phase, the proposed pseudotriggered mechanism is expected to remain robust below the JT ordering temperature (known to be above room temperature~\cite{Matsushita2005,Yu2024RbI}), as the symmetry arguments and mode couplings are intrinsic to the crystal structure. Incorporating finite-temperature effects e.g., via \textit{ab-initio} molecular dynamics or anharmonic phonon calculations could be a valuable extension of this work.}

%Cs$_2\rm{Au}_2$I$_6$ has been established as a potential absorber material for ultrahigh efficiency Pb‐free photovoltaic applications. While the nonpolar $I4/mmm$ phase of this material shows high absorption, and i
%While the nonpolar $I4/mmm$ phase of Cs$_2\rm{Au}_2$I$_6$ has been established as a potential absorber material for ultrahigh efficiency photovoltaic (PV) applications, polar materials with an intrinsic in-built potential are sought for high efficiency ferroelectric-photovoltaic (FE-PV) device applications.  In FE-PVs, the photovoltage is not limited by the bandgap of the ferroelectric material, and the recombination rates, which can impede PV efficiency, are minimized. Therefore, 

\section{Acknowledgements}
We acknowledge the Leverhulme Trust for a research project grant (Grant No. RPG-2020-206). This work used the Hamilton HPC Service of Durham University. We thank Emma E. McCabe for useful discussions.

\bibliography{bib_paper}

\clearpage
\newpage

\onecolumngrid

\section*{SUPPLEMENTAL MATERIAL}
\input{./SM_merge_contents}

 \end{document}

%% file: SM_merge_contents.tex
\setcounter{page}{1}
\setcounter{figure}{0}
\setcounter{table}{0}
\setcounter{section}{0}
\renewcommand{\thepage}{S\arabic{page}}
\renewcommand{\thesection}{S\arabic{section}}
\renewcommand{\thetable}{S\arabic{table}}
\renewcommand{\thefigure}{S\arabic{figure}}
\newcounter{SIfig}
\renewcommand{\theSIfig}{S\arabic{SIfig}}

\begin{table}[ht!]
  \setlength{\tabcolsep}{6.0pt}
  \caption{DFT-optimized lattice parameters of the ground state (G.S.) structures and the corresponding mode amplitudes ($A_p$ values in ISODISTORT~\cite{Campbell1,isodistort2}) of $A_2\rm{Au}_2$$X_6$ ($A$: Cs, Rb; $X$: Cl, Br, I)  double perovskites. All the structures have $\alpha$ = $\gamma$ = 90\degree. Goldschmidt tolerance factors ($\tau$) and available experimental lattice parameters are given in parentheses.}
  \centering
  \begin{tabular}{|c |c |c | c| c| c| c| c|}
    \hline
    \hline\rule{0pt}{1.0\normalbaselineskip}
%    \multicolumn{2}{c|}{$A_2\rm{Au}_2X_6$}&\multicolumn{2}{c|}{Strain modes}&\multicolumn{2}{c|}{Displacive modes}\\\cline{1-6}    \rule{0pt}{1.2\normalbaselineskip}
     Gold &G.S.  &  \multicolumn{4}{c|}{Relaxed lattice parameters} & Cell basis & Relaxed mode  \\ \cline{3-6} \rule{0pt}{1.0\normalbaselineskip}
    halide &structure& $a$ ({\AA}) & $b$ ({\AA}) &$c$ ({\AA}) & $\beta$ (\degree) &w.r.t. cubic parent &amplitudes ({\AA}) \\ \hline \rule{0pt}{1.0\normalbaselineskip}
     Cs$_2$Au$_2$Cl$_6$& $I4/mmm$ & 7.456 & 7.456 & 11.075 & 90 &\{(1,1,0),(-1,1,0),(0,0,2)\}& $Q_{\Gamma^+_1(a)}$ = 0.086 \\\rule{0pt}{1.0\normalbaselineskip}
    ($\tau$ = 0.920)&&(7.500)$^*$&(7.500)$^*$&(10.884)$^*$&(90)$^*$&&$Q_{\Gamma^+_3(a, 0)}$ = 0.042\\\rule{0pt}{1.0\normalbaselineskip}
     &&&&&&&$Q_{R^-_2(a)}$ = 0.088\\\rule{0pt}{1.0\normalbaselineskip}
     &&&&&&&$Q_{R^-_3(a, 0)}$ = 0.606\\ \hline \rule{0pt}{1.0\normalbaselineskip}
     Cs$_2$Au$_2$Br$_6$& $I4/mmm$ & 7.768 & 7.768 & 11.550 & 90 & \{(1,1,0),(-1,1,0),(0,0,2)\}&$Q_{\Gamma^+_1(a)}$ = 0.070 \\\rule{0pt}{1.0\normalbaselineskip}
     ($\tau$ = 0.910)&&(7.787)$^\intercal$&(7.787)$^\intercal$&(11.380)$^\intercal$&(90)$^\intercal$&&$Q_{\Gamma^+_3(a, 0)}$ = 0.043\\\rule{0pt}{1.0\normalbaselineskip}
     &&&&&&&$Q_{R^-_2(a)}$ = 0.051\\\rule{0pt}{1.0\normalbaselineskip}
     &&&&&&&$Q_{R^-_3(a, 0)}$ = 0.558\\ \hline \rule{0pt}{1.0\normalbaselineskip}
     Cs$_2$Au$_2$I$_6$& $I4/mmm$ & 8.292 & 8.292 & 12.275 & 90 & \{(1,1,0),(-1,1,0),(0,0,2)\}&$Q_{\Gamma^+_1(a)}$ = 0.065 \\\rule{0pt}{1.0\normalbaselineskip}
     ($\tau$ = 0.895)&&(	
8.285)$^\divideontimes$&(	
8.285)$^\divideontimes$&(12.084)$^\divideontimes$&(90)$^\divideontimes$&&$Q_{\Gamma^+_3(a, 0)}$ = 0.039 \\ \rule{0pt}{1.0\normalbaselineskip}
     &&&&&&&$Q_{R^-_2(a)}$ = 0.036\\\rule{0pt}{1.0\normalbaselineskip}
     &&&&&&&$Q_{R^-_3(a, 0)}$ = 0.563\\ \hline \rule{0pt}{1.0\normalbaselineskip}
     Rb$_2$Au$_2$Cl$_6$& $I4/mmm$ & 7.362 & 7.362 & 10.916 & 90 & {(1,1,0),(-1,1,0),(0,0,2)}&$Q_{\Gamma^+_1(a)}$ = 0.076 \\\rule{0pt}{1.0\normalbaselineskip}
     ($\tau$ = 0.880)&&&&&&&$Q_{\Gamma^+_3(a, 0)}$ = 0.041\\\rule{0pt}{1.0\normalbaselineskip}
     &&&&&&&$Q_{R^-_2(a)}$ = 0.072\\\rule{0pt}{1.0\normalbaselineskip}
     &&&&&&&$Q_{R^-_3(a, 0)}$ = 0.546\\ \hline \rule{0pt}{1.0\normalbaselineskip}
     Rb$_2$Au$_2$Br$_6$& $C2/m$ & 12.893 & 7.420 & 8.170 & 118.322 & \{(2,1,-1),(0,1,1),(0,-1,1)\}&$Q_{\Gamma^+_1(a)}$ = 0.078 \\\rule{0pt}{1.0\normalbaselineskip}
     ($\tau$ = 0.872)&&(12.689)$\dagger$&(7.243)$\dagger$&(8.520)$\dagger$&(119.950)$\dagger$&&$Q_{\Gamma^+_3(a, 1.732a)}$ = 0.037\\\rule{0pt}{1.0\normalbaselineskip}
     &&&&&&&$Q_{\Gamma^+_5(a, b, -a)}$ = 0.157\\\rule{0pt}{1.0\normalbaselineskip}
     &&&&&&&$Q_{R^-_2(a)}$ = 0.057\\\rule{0pt}{1.0\normalbaselineskip}
     &&&&&&&$Q_{R^-_3(a, 1.732a)}$ = 0.521\\\rule{0pt}{1.0\normalbaselineskip}
     &&&&&&&$Q_{R^-_4(a, b, -a)}$ = 0.293\\
     \rule{0pt}{1.0\normalbaselineskip}
     &&&&&&&$Q_{R^-_5(a, 0, a)}$ = 0.691\\
     \hline \rule{0pt}{1.0\normalbaselineskip}
     Rb$_2$Au$_2$I$_6$& $C2/m$ & 13.587 & 8.006 & 8.680 & 117.027 & \{(2,1,-1),(0,1,1),(0,-1,1)\}&$Q_{\Gamma^+_1(a)}$ = 0.076 \\\rule{0pt}{1.0\normalbaselineskip}
     ($\tau$ = 0.859)&&(13.485)$\ddagger$&(7.970)$\ddagger$&(8.762)$\ddagger$&(118.042)$\ddagger$&&$Q_{\Gamma^+_3(a, 1.732a)}$ = 0.038\\\rule{0pt}{1.0\normalbaselineskip}
     &&&&&&&$Q_{\Gamma^+_5(a, b, -a)}$ = 0.165\\\rule{0pt}{1.0\normalbaselineskip}
     &&&&&&&$Q_{R^-_2(a)}$ = 0.040\\\rule{0pt}{1.0\normalbaselineskip}
     &&&&&&&$Q_{R^-_3(a, 1.732a)}$ = 0.534\\\rule{0pt}{1.0\normalbaselineskip}
     &&&&&&&$Q_{R^-_4(a, b, -a)}$ = 0.376\\
     \rule{0pt}{1.0\normalbaselineskip}
     &&&&&&&$Q_{R^-_5(a, 0, a)}$ = 0.793\\
    \hline 
     \hline 
    \multicolumn{8}{l}{$*$ denotes the experimental values measured in Ref.~\cite{Matsushita1998}.} \\
    \multicolumn{8}{l}{$\intercal$ denotes the experimental values measured in Ref.~\cite{Scott2012}.} \\
    \multicolumn{8}{l}{$\divideontimes$ denotes the experimental values measured in Ref.~\cite{Scott2012}.} \\
    \multicolumn{8}{l}{$\dagger$ denotes the experimental values measured in Ref.~\cite{Strahle1979}.} \\
    \multicolumn{8}{l}{$\ddagger$ denotes the experimental values measured in Ref.~\cite{Matsushita2005}.} \\
%     \hline \rule{0pt}{1.0\normalbaselineskip}
     
%     \hline  
\end{tabular}
\label{structure_all}
\end{table}

\begin{figure}[b]
\includegraphics[scale=1.5]{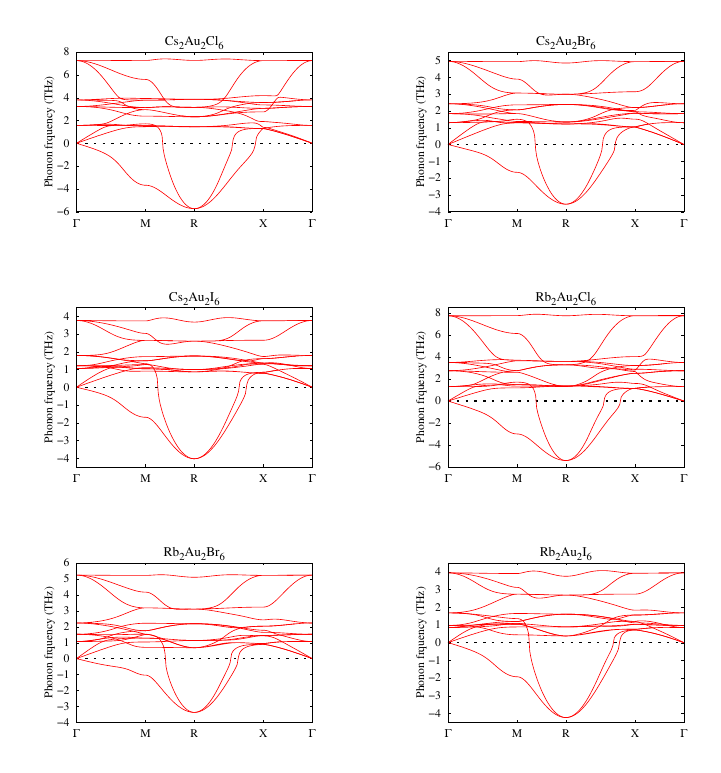}
\caption{Phonon spectra of the high-symmetry cubic perovskite phase of $A$AuI$_3$ gold halide perovskites ($A$: Cs, Rb; $X$: Cl, Br, I).} 
\label{phonon_all}
\end{figure} 

\begin{figure}[b]
\includegraphics[scale=1.5]{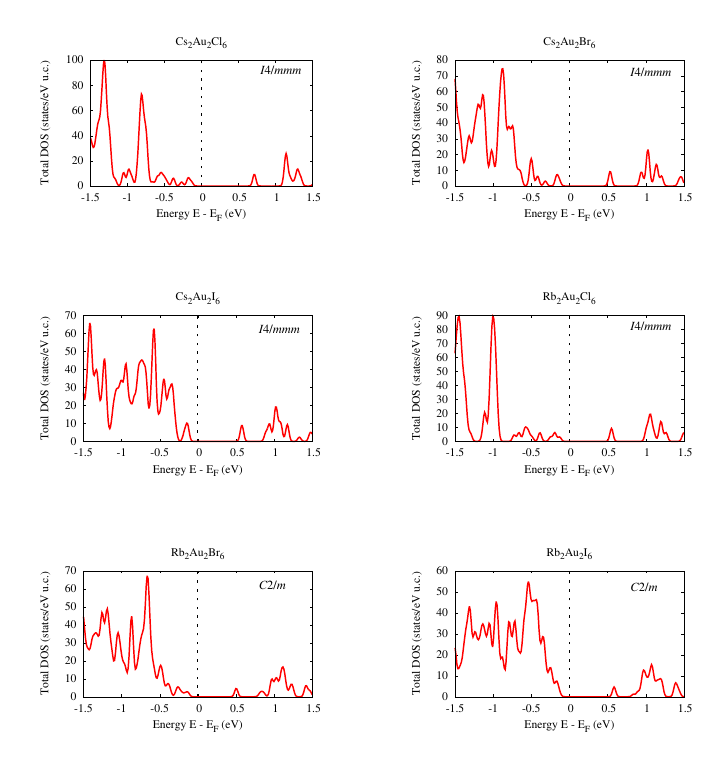}
\caption{Total density of states (DOS) calculated for the ground state structures of $A_2\rm{Au}_2$$X_6$ ($A$: Cs, Rb; $X$: Cl, Br, I) double perovskites.} 
\label{DOS_all}
\end{figure} 

\begin{table}[ht!]
  \setlength{\tabcolsep}{6.0pt}
  \caption{DFT-optimized lattice parameters of the previously unreported polar $I4mm$ phase of Rb$_2\rm{Au}_2$I$_6$ and the corresponding mode amplitudes ($A_p$ values in ISODISTORT~\cite{Campbell1,isodistort2}).}
  \centering
  \begin{tabular}{|c |c |c | c| c| c| c| c|}
    \hline
    \hline\rule{0pt}{1.0\normalbaselineskip}
    $a$ ({\AA}) & $b$ ({\AA}) &$c$ ({\AA}) & $\alpha$ (\degree) &$\beta$ (\degree) &$\gamma$ (\degree) & Cell basis w.r.t. cubic parent &Mode amplitudes ({\AA}) \\ \hline \rule{0pt}{1.0\normalbaselineskip}
    8.268 & 8.268 & 12.353 & 90 &90 & 90 &\{(1,1,0),(-1,1,0),(0,0,2)\}& $Q_{\Gamma^+_1(a)}$ = 0.072 \\\rule{0pt}{1.0\normalbaselineskip}
    &&&&&&&$Q_{\Gamma^+_3(a, 0)}$ = 0.047\\\rule{0pt}{1.0\normalbaselineskip}
     &&&&&&&$Q_{\Gamma^-_4(0, 0, a)}$ = 0.878\\\rule{0pt}{1.0\normalbaselineskip}
     &&&&&&&$Q_{R^+_5(a, 0 , 0)}$ = 0.016\\\rule{0pt}{1.0\normalbaselineskip}
     &&&&&&&$Q_{R^-_2(a)}$ = 0.023\\\rule{0pt}{1.0\normalbaselineskip}
     &&&&&&&$Q_{R^-_3(a, 0)}$ = 0.565\\ \hline      \hline  
\end{tabular}
\label{structure_I4mm}
\end{table}

\begin{table}[ht!]
  \setlength{\tabcolsep}{6.0pt}
  \caption{\ud{Components of the static Dielectric permittivity tensor $\epsilon^\infty_{ij} ~(i, j = x, y, z)$ calculated for the polar $I4mm$ phase of Rb$_2\rm{Au}_2$I$_6$. $\epsilon^{\infty}_{\text{el}}$ and $\epsilon^{0}_{\text{ion}}$ represent the electronic and ionic contributions, respectively.}}
  \label{tab-diel}
  \begin{tabular}{|c c c c c c c c c c|}
    \hline
     $\epsilon^{\infty}$&$xx$&$xy$&$xz$&$yx$&$yy$&$yz$&$zx$&$zy$&$zz$ \\
    \hline
    $\epsilon^{\infty}_{\text{el}}$&18.127493&0&0&0&18.127493&0&0&0&4.584249\\[0.1cm]
    $\epsilon^{0}_{\text{ion}}$&26.473820&0&0&0&26.473820&0&0&0&8.601037\\[0.1cm]
    \hline
\end{tabular}
\end{table}

\begin{figure}[b]
\includegraphics[scale=1.5]{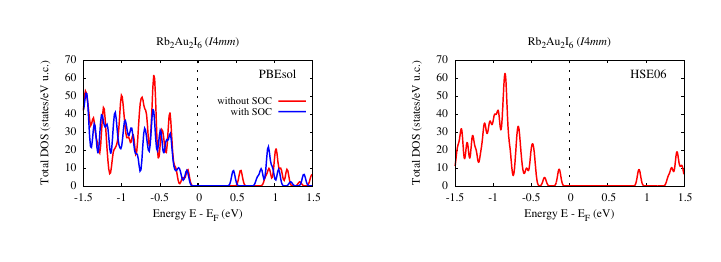}
\caption{\ud{Total density of states (DOS) computed for the $I4mm$ structure of Rb$_2\rm{Au}_2$I$_6$ using PBEsol (left) and HSE06 hybrid (right) functionals, showing the insulating nature of this novel polar phase. Effect of spin-orbit coupling (SOC) is considered only with the PBEsol functional.}} 
\label{DOS_polar}
\end{figure}

%\begin{figure}[b]
%\includegraphics[scale=0.8]{Fig4_SM.png}
%\caption{\ud{Total density of states (DOS) computed for the $I4mm$ structure of Rb$_2\rm{Au}_2$I$_6$ using PBEsol (left) and HSE06 hybrid (right) functionals, showing the insulating nature of this novel polar phase. Effect of spin-orbit coupling (SOC) is considered only with the PBEsol functional.}} 
%\label{HSE_SOC}
%\end{figure}

\begin{figure}[b]
\includegraphics[scale=1.5]{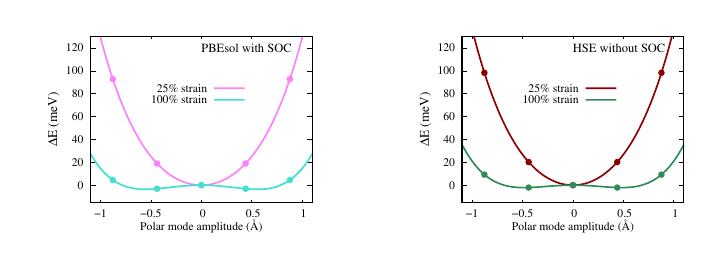}
\caption{\ud{Energy landscape of the $I4mm$ phase as a function of the polar mode amplitude at two different percentage values of the strain modes (both hydrostatic and tetragonal) produced by the Jahn-Teller distortion in the $I4/mmm$ phase, calculated using PBEsol functional+SOC (left) and HSE06 hybrid functional (right). The calculations exclude the contributions from secondary breathing and antipolar modes. Since no structural relaxations were performed with either SOC or the HSE06 functional, the relaxed strain and polar mode amplitudes correspond to the PBEsol-optimized structure. The similarity of these results with those obtained using the PBEsol functional in Fig.~4 of the main text further demonstrates the robustness of the proposed pseudo-triggered mechanism.}} 
\label{HSE_SOC}
\end{figure}